\documentclass[allclo]{FBSart}
\usepackage{amsfonts}
\usepackage{amssymb}
\usepackage{graphicx} 

\newcommand{\sfrac}[2]{\mbox{\footnotesize $\displaystyle \frac{#1}{#2}$}}

\title{Regarding proton form factors} 

\author{J.\,C.\,R.\ Bloch,\instnr{1} A.\ Krassnigg\instnr{2} and C.\,D.\
Roberts\instnr{2}} 

\instlist{DFG Research Center ``Mathematics for Key Technologies,''
c/o Weierstrass Institute for Applied Analysis and Stochastics,
Mohrenstrasse 39, D-10117 Berlin , Germany
\and 
Physics Division, Argonne National Laboratory, Argonne, IL
60439-4843, USA}

\runningauthor{J.\,C.\,R.\ Bloch, et al.}
\runningtitle{Regarding proton form factors}
\begin{document}

\maketitle
\begin{abstract}
The proton's elastic electromagnetic form factors are calculated using an
\textit{Ansatz} for the nucleon's Poincar\'e covariant Faddeev amplitude that
only retains scalar diquark correlations.  A spectator approximation is
employed for the current.  On the domain of $q^2$ accessible in modern
precision experiments these form factors are a sensitive probe of
nonperturbative strong interaction dynamics.  The ratio of Pauli and Dirac
form factors can provide realistic constraints on models of the nucleon and
thereby assist in developing an understanding of nucleon structure.

\received{20/June/03}
\end{abstract}

\section{Introduction}
The pion's elastic electromagnetic form factor is accessible via a properly
constructed six-point quark Schwinger function, a fact exploited with modest
success in lattice-QCD simulations, e.g., Refs.\
\cite{latticepion1,draper,laermannpion}.  This Schwinger function is also the
basis for continuum studies, among which those employing the Dyson-Schwinger
equations (DSEs) \cite{revbasti,revreinhard,revpieter} are efficacious
\cite{pieterpion}.  The proton's form factors are accessible through an
analogous eight-point quark Schwinger function, which is the starting point
for lattice simulations, e.g. Refs.\ \cite{latticeform1,latticeform}.  The
fruitful extension of DSE methods to the calculation of this Schwinger
function is a contemporary goal.  As we will explain, a simple truncation
that corresponds to a spectator approximation is currently in widespread use
\cite{bloch1,bloch2,oettel1}.  Manifest covariance is a strength of this
approach for it has long been apparent that in order to obtain an internally
consistent understanding of proton form factor data at spacelike momentum
transfers $q^2 \gtrsim M^2$, where $M$ is the proton's mass, a Poincar\'e
covariant description of the scattering process is necessary
\cite{coester,revcoester}.  This has recently been reemphasised in the
context of constituent-quark models \cite{salme,plessas,millerfrank}.

The same interaction which describes the structure and properties of
colour-singlet mesons also generates a quark-quark (diquark) correlation in
the colour antitriplet ($\bar 3$) channel \cite{justindq,pieterdq}.  Such
correlations have recently been observed in simulations of lattice-QCD
\cite{latticedq}.  While diquarks do not survive as asymptotic states
\cite{truncscheme,reinharddq,detmold}; viz., they do not appear in the strong
interaction spectrum, the existence of strong quark-quark correlations
provides a foundation for viewing the nucleon as a quark-diquark composite.
This picture can be realised via a Poincar\'e covariant Faddeev equation
\cite{regfe}, in which two quarks are always correlated as a $\bar
3$-diquark, and binding in the nucleon is effected by the iterated exchange
of roles between the dormant and diquark-participant quarks, and through the
action of a pion ``cloud'' \cite{NpiN}.

Upon solving the Faddeev equation one obtains the nucleon's mass, and also
its Faddeev amplitude which is a valuable intuitive tool.  It is noteworthy
that even a rudimentary covariant Faddeev equation model, based on the
presence of diquark correlations within the nucleon, yields a matrix-valued
amplitude that, in the nucleon's rest frame, corresponds to a relativistic
wave function with a material lower component; i.e., a wave function with
``$p$-wave,'' and, indeed, ``$d$-wave'' correlations, too \cite{oettel2}.
Nonzero quark orbital angular momentum in the nucleon is a straightforward
outcome of a Poincar\'e covariant description.

While some issues remain unresolved \cite{brash,arrington}, contemporary data
\cite{jones,roygayou,gayou} suggest that a single dipole mass cannot
simultaneously characterise the $Q^2$-dependence of both the proton's
electric and magnetic form factors.  This possibility was evident in the
Faddeev-amplitude-based calculations of Ref.\ \cite{bloch1}, as emphasised in
Ref.\ \cite{bloch2}.  Moreover, it can be inferred from Ref.\ \cite{ralston}
that this experimental result is an essential consequence of a
nonperturbative and Poincar\'e covariant representation of the proton as a
bound state.  This may be exemplified through the role played by pseudovector
components of the pion's Bethe-Salpeter amplitude, which are connected with
the presence of quark orbital angular momentum in the pion.  These
pseudovector amplitudes are necessarily nonzero \cite{mrt98} and responsible
for the large-$Q^2$ behaviour of the electromagnetic pion form factor
\cite{jackson,mrpion}.

Herein we calculate the proton's elastic electromagnetic form factors, using
a product \textit{Ansatz} for the proton's Faddeev amplitude and a spectator
approximation to describe elastic electromagnetic scattering from the
nucleon.  We describe the model in Sect.\ \ref{sec:model}, and present and
discuss the results in Sect.\ \ref{sec:results}.  Section \ref{sec:epilogue}
is an epilogue.  The study furnishes a means by which we may explore and
illustrate the points outlined above.  It will become apparent that existing
precision data on the form factor ratios
$F_2^p(q^2)/F_1^p(q^2)$ and 
$G_E^p(q^2)/G_M^p(q^2)$ 
are a sensitive probe of nonperturbative aspects of the proton's structure.

\section{Model Elements}
\label{sec:model}
\subsection{Dressed quarks} 
\label{subsec:Sp}
There are three primary elements of our model and to begin with its
specification we note that quarks within bound states are described by a
dressed propagator\footnote{We employ a Euclidean metric wherewith the scalar
product of two four vectors is \mbox{$a\cdot b=\sum_{i=1}^4 a_i b_i$}, and
Hermitian Dirac-$\gamma$ matrices that obey \mbox{$\{\gamma_\mu,\gamma_\nu\}
= 2\, \delta_{\mu\nu}$}.}
\begin{equation}
\label{SpAB}
S(p)  =  - i \gamma\cdot p \, \sigma_V(p^2) + \sigma_S(p^2)
\;  =  \; \frac{1}{i \gamma\cdot p \, A(p^2) + B(p^2)}
\;  =  \; \frac{Z(p^2)}{i \gamma\cdot p + M(p^2)}\,.
\end{equation}
It is a longstanding, model-independent DSE prediction that the wave function
renormalisation $Z(p^2)$ and mass function $M(p^2)$ exhibit significant
momentum dependence for $p^2\lesssim 2\,$GeV$^2$ whose origin is
nonperturbative \cite{revbasti,revreinhard,revpieter}.  This behaviour was
recently verified in numerical simulations of quenched QCD \cite{bowman}, and
the connection between this and the full theory is analysed in Ref.\
\cite{mandar}.

The mass function is enhanced at infrared momenta, a feature that is an
essential consequence of the dynamical chiral symmetry breaking (DCSB)
mechanism.  It is also the origin of the constituent-quark mass.  With
increasing spacelike $p^2$, on the other hand, the mass function evolves to
reproduce the asymptotic behaviour familiar from perturbative analyses and
that behaviour is manifest for $p^2 \gtrsim 10\,$GeV$^2$ \cite{mr97}.

While numerical solutions for the dressed-quark propagator are readily
obtained from a model of QCD's gap equation, the utility of an algebraic form
for $S(p)$ when calculations require the evaluation of numerous
multidimensional integrals is self-evident.  An efficacious parametrisation,
which exhibits the features described above and has been used extensively
\cite{revbasti,revreinhard,revpieter}, is expressed via
\begin{eqnarray} 
\bar\sigma_S(x) & =&  2\,\bar m \,{\cal F}(2 (x+\bar m^2)) + {\cal
F}(b_1 x) \,{\cal F}(b_3 x) \,  
\left[b_0 + b_2 {\cal F}(\epsilon x)\right]\,,\label{ssm} \\ 
\label{svm} \bar\sigma_V(x) & = & \frac{1}{x+\bar m^2}\, \left[ 1 - {\cal F}(2 
(x+\bar m^2))\right]\,, 
\end{eqnarray}
with $x=p^2/\lambda^2$, $\bar m$ = $m/\lambda$, 
\begin{equation}
\label{defcalF}
{\cal F}(x)= \frac{1-\mbox{\rm e}^{-x}}{x}  \,, 
\end{equation}
$\bar\sigma_S(x) = \lambda\,\sigma_S(p^2)$ and $\bar\sigma_V(x) =
\lambda^2\,\sigma_V(p^2)$.  The mass-scale, $\lambda=0.566\,$GeV, and
parameter values\footnote{$\epsilon=10^{-4}$ in Eq.\ (\ref{ssm}) acts only to
decouple the large- and intermediate-$p^2$ domains.}
\begin{equation} 
\label{tableA} 
\begin{array}{ccccc} 
   \bar m& b_0 & b_1 & b_2 & b_3 \\\hline 
   0.00897 & 0.131 & 2.90 & 0.603 & 0.185 
\end{array}\;, 
\end{equation} 
were fixed in a least-squares fit to light-meson observables \cite{mark}.
The dimensionless $u=d$ current-quark mass in Eq.~(\ref{tableA}) corresponds
to
\begin{equation} 
m=5.1\,{\rm MeV}\,. 
\end{equation} 

The parametrisation yields a Euclidean constituent-quark mass
\begin{equation} 
\label{MEq} M_{u,d}^E = 0.33\,{\rm GeV}, 
\end{equation} 
defined as the solution of $p^2=M^2(p^2)$ \cite{mr97}, whose magnitude is
typical of that employed in constituent-quark models \cite{simon}.  This is
an expression of DCSB, as is the vacuum quark condensate
\begin{equation} 
-\langle \bar qq \rangle_0^{1\,{\rm GeV}^2} = \lambda^3\,\frac{3}{4\pi^2}\, 
\frac{b_0}{b_1\,b_3}\,\ln\frac{1\,{\rm GeV}^2}{\Lambda_{\rm QCD}^2} = 
(0.221\,{\rm GeV})^3\,, 
\end{equation}  
$\Lambda_{\rm QCD}=0.2\,$GeV.  The condensate is calculated directly from its
gauge invariant definition \cite{mrt98} after making allowance for the fact
that Eqs.\ (\ref{ssm}), (\ref{svm}) yield a chiral-limit quark mass function
with anomalous dimension $\gamma_m = 1$.  This omission of the additional
$\ln( p^2/\Lambda_{\rm QCD}^2)$-suppression that is characteristic of QCD is
merely a practical simplification.

\subsection{Product \textit{Ansatz} for the Faddeev amplitude}
We represent the proton as a composite of a dressed-quark and nonpointlike,
Lorentz-scalar quark-quark correlation (diquark), and exhibit this via a
product \textit{Ansatz} for the Faddeev amplitude
\begin{eqnarray} 
\Psi_3^{0^+}(p_i,\alpha_i,\tau_i) &=&
[\Gamma^{0^+}(\frac{1}{2}p_{[12]};K)]_{\alpha_1  
\alpha_2}^{\tau_1 \tau_2}\, \Delta^{0^+}(K) \,[{\cal S}(\ell;P)
u(P)]_{\alpha_3}^{\tau_3}\,,
\label{calS} 
\end{eqnarray}
wherein $(p_i,\alpha_i,\tau_i)$ are the momentum, spin and isospin labels of
the quarks constituting the nucleon; the spinor satisfies
\begin{equation}
\bar u(P)\, (i\gamma\cdot P + M) = 0 = (i\gamma\cdot P + M)\, u(P) \,,
\end{equation}
with $P=p_1+p_2+p_3$ the nucleon's total momentum, and it is also a spinor in
isospin space with $\varphi_+= {\rm col}(1,0)$ for the proton and $\varphi_-=
{\rm col}(0,1)$ for the neutron; and $K= p_1+p_2=: p_{\{12\}}$, $p_{[12]}= p_1 -
p_2$, $\ell := (-p_{\{12\}} + 2 p_3)/3$.

In Eq.\ (\ref{calS}), $\Delta^{0^+}(K)$ is a pseudoparticle propagator for
the scalar diquark formed from quarks $1$ and $2$, and $\Gamma^{0^+}\!$ is a
Bethe-Salpeter-like amplitude describing their relative momentum correlation.
These functions can be obtained from an analysis of the quark-quark
scattering matrix, as explained in Ref.\ \cite{NpiN}.  However, we have
already chosen to simplify our calculations by parametrising $S(p)$, and
hence we follow Refs.\ \cite{bloch1,bloch2} and also employ that expedient
herein, using
\begin{eqnarray} 
\label{Delta0p}
\Delta^{0^+}(K) & = & \frac{1}{m_{0^+}^2}\,{\cal F}(K^2/\omega_{0^+}^2)\,,\\ 
\label{Gamma0p} \Gamma^{0^+}(k;K) &=& \frac{1}{{\cal N}^{0^+}} \, 
H\,C i\gamma_5\, i\tau_2\, {\cal F}(k^2/\omega_{0^+}^2) \,, 
\end{eqnarray} 
with ${\cal F}$ defined in Eq.\ (\ref{defcalF}), $C=\gamma_2\gamma_4$ the
charge conjugation matrix, $\tau_2$ the $2\times 2$ Pauli isospin matrix, and
$ (H^{c_3})_{c_1 c_2} = \epsilon_{c_1 c_2 c_3}$, $c_{1,2,3}=1,2,3$,
describing the completely antisymmetric colour structure of a $\bar 3$
diquark.\footnote{In Eq.\ (\protect\ref{Gamma0p}), ${\cal N}^{0^+}$ is a
calculated normalisation constant which ensures that a $(ud)$-diquark has
electric charge fraction $(1/3)$ for $K^2=-m_{0^+}^2$.  ${\cal N}_\Psi$, to
appear in Eq.\ (\protect\ref{SAnsatz}), is analogous: it is the calculated
normalisation constant that ensures the proton has unit charge.}  The
parameters are a width, $\omega_{0^+}$, and a pseudoparticle mass, $m_{0^+}$,
which have ready physical interpretations: the length
$l_{0^+}=1/\omega_{0^+}$ is a measure of the mean separation between the
quarks in the scalar diquark; and the distance $\lambda_{0^+}= 1/m_{0^+}$
represents the range over which a diquark correlation in this channel can
persist inside the nucleon.

With the elements described hitherto it is possible to derive a Poincar\'e
covariant Faddeev equation whose solution yields ${\cal S}$, a $4\times 4$
Dirac matrix that describes the relative quark-diquark momentum correlation.
The complete nucleon amplitude then follows:
\begin{equation}
\label{PsiT}
\Psi = \Psi_1 + \Psi_2 + \Psi_3 \,, 
\end{equation} 
where the subscript identifies the dormant quark and, e.g., $\Psi_{1,2}$ are
obtained from $\Psi_3$ by a uniform, cyclic permutation of all the quark
labels.  The general form of ${\cal S}$ is discussed at length in Ref.\
\cite{oettel2} with the conclusion that the positive energy solution can be
written
\begin{equation} 
{\cal S}(\ell;P) = f_1(\ell;P)\,I_{\rm D} + \frac{1}{M}\left(i\gamma\cdot \ell 
- \ell \cdot \hat P\, I_{\rm D}\right)\,f_2(\ell;P)\,, 
\end{equation}
where $(I_{\rm D})_{rs}= \delta_{rs}$, $\hat P^2= - 1$.  In the nucleon's
rest frame, $f_{1,2}$, respectively, describe the upper, lower component of
the spinor amplitude ${\cal S}(\ell;P)\, u(P)$.

Again, calculations are simplified if one employs an algebraic
parametrisation of ${\cal S}$, and the form \cite{nedm}
\begin{equation}
\label{SAnsatz}
{\cal S}(\ell;P) = \frac{1}{{\cal N}_\Psi} \, {\cal
F}(\ell^2/\omega^2_{q\{qq\}}) \, \left[ I_{\rm D} - \frac{\mbox{\sc r}}{M}
\left( i\gamma\cdot \ell - \ell \cdot \hat P\, I_{\rm D}\right) \right]
\end{equation}
is efficacious.  In writing this one exploits results of the Faddeev equation
calculations reported in Refs.\ \cite{NpiN,oettel2}, which establish the
fidelity of the approximations $f_2(\ell;P) \approx \mbox{\sc r} f_1(\ell;P)$
and $f_1(\ell;P) \approx f_1(\ell^2;P^2)$.  The \textit{Ansatz} involves two
parameters: a width $\omega_{q\{qq\}}$ and ratio {\sc r}.  The former can be
associated with a length-scale $l_{q\{qq\}} = 1/\omega_{q\{qq\}}$, which
measures the quark-diquark separation.  The latter gauges the importance of
the lower component of the positive energy nucleon's spinor amplitude.  Its
magnitude increases with increasing {\sc r}.  (The strength of the lower
component of the nucleon's Faddeev \textit{wave function} is determined by
{\sc r} but does not vanish for $\mbox{\sc r} = 0$.)  In realistic Faddeev
equation studies of the nucleon $l_{q\{qq\}} > l_{0^+}/2 \sim 0.2\,$fm and
$\mbox{\sc r} \sim 0.5$ \cite{NpiN}.

\subsection{Dressed-quark-photon coupling}
A calculation of the electromagnetic interaction of a composite particle
cannot proceed without an understanding of the coupling between the photon
and the bound state's constituents.  If those constituents are dressed then
the coupling is not pointlike.  Indeed, it is readily apparent that with
quarks dressed as described in Sec.~\ref{subsec:Sp}, only a
dressed-quark-photon vertex, $\Gamma_\mu$, can satisfy the vector
Ward-Takahashi identity:
\begin{equation}
\label{vwti}
q_\mu \, i\Gamma_\mu(\ell_1,\ell_2) = S^{-1}(\ell_1) -
S^{-1}(\ell_2)\,,
\end{equation}
where $q=\ell_1-\ell_2$ is the photon momentum flowing into the vertex.  This
is illustrated with particular emphasis in
Refs.~\cite{mrpion,anomalyBando,anomalyRoberts1,anomalyAlkofer,anomalyKlabucar,anomalyRoberts2,anomalyTandy},
which consider effects associated with the Abelian anomaly.  The
Ward-Takahashi identity is only one of many constraints that apply to
$\Gamma_\mu$ in a renormalisable quantum field theory and these have been
explored extensively in Refs.\ \cite{ayse1,ayse2}.

The dressed-quark-photon vertex, a three-point Schwinger function, can be
obtained by solving an inhomogeneous Bethe-Salpeter equation.  This was the
procedure adopted in the DSE calculation \cite{pieterpion} that successfully
predicted the electromagnetic pion form factor \cite{Volmer:2000ek}.  For our
purposes, however, it is enough to follow Ref.\ \cite{mark} and employ the
algebraic parametrisation \cite{bc80}
\begin{equation}
\label{bcvtx}
i\Gamma_\mu(\ell_1,\ell_2)  =  
i\Sigma_A(\ell_1^2,\ell_2^2)\,\gamma_\mu +
(\ell_1+\ell_2)_\mu \left[\sfrac{1}{2}i\gamma\cdot (\ell_1+\ell_2) \,
\Delta_A(\ell_1^2,\ell_2^2) + \Delta_B(\ell_1^2,\ell_2^2)\right];
\end{equation}
with 
\begin{equation}
\Sigma_F(\ell_1^2,\ell_2^2) = \sfrac{1}{2}\,[F(\ell_1^2)+F(\ell_2^2)]\,,\;
\Delta_F(\ell_1^2,\ell_2^2) =
\frac{F(\ell_1^2)-F(\ell_2^2)}{\ell_1^2-\ell_2^2}\,,
\label{DeltaF}
\end{equation}
where $F= A, B$; i.e., the scalar functions in Eq.~(\ref{SpAB}).  It is
critical that $\Gamma_\mu$ in Eq.\ (\ref{bcvtx}) satisfies Eq.\ (\ref{vwti})
and very useful that it is completely determined by the dressed-quark
propagator.  Improvements to this \textit{Ansatz} modify results by $\lesssim
10\,$\%, as illustrated, e.g., in Refs. \cite{pichowsky,piloop}.

Equation (\ref{bcvtx}) entails that dressed-quarks do not respond as point
particles to low momentum transfer probes.  This observation qualitatively
supports an assumption employed in some relativistic constituent quark models
\cite{coester,salme,salme2}.  An unambiguous quantitative connection is
difficult because the definition of constituent-quark degrees of freedom
depends on a model's formulation.  It may nevertheless be worth noting that
quark dressing disappears with increasing spacelike $q^2$ in QCD.  Hence, in
the ultraviolet, the dressed-quark's Dirac form factor must approach one
(with only $\ln q^2$ corrections) and its Pauli form factor must vanish;
viz., the interaction becomes pointlike in this limit.  With the parameter
values in Eq.\ (\ref{tableA}), this evolution of the Dirac and Pauli form
factors may be characterised by monopole ranges $r_1 \sim 0.25\,$fm and $r_2
\sim 0.35\,$fm, respectively.

\subsection{Commentary}
We have completely specified a covariant model of the nucleon as a bound
state of a dressed-quark and nonpointlike scalar quark-quark correlation.
This algebraic \textit{Ansatz} has four parameters: $m_{0^+}$ and
$\omega_{0^+}$ introduced in Eqs.\ (\ref{Delta0p}), (\ref{Gamma0p}) to
characterise the diquark; and $\omega_{q\{qq\}}$ and {\sc r} in Eq.\
(\ref{SAnsatz}), which express prominent features of the nucleon's spinor.
The dressed-quark propagator and dressed-quark-photon vertex are fixed.

In contemplating such a model one may ask whether it \textit{can} supply an
accurate description of the nucleon's electromagnetic form factors.
\textit{A priori}, the answer is unknown but it is supplied by
straightforward calculations.

A more important question, however, is whether the model \textit{should} be
accurate.  In this case the answer is \textit{no}.  Reference \cite{NpiN}
emphasises that no picture of the nucleon is veracious if it neglects
axial-vector diquark correlations and the nucleon's pion ``cloud.''  Thus our
simple model must be incomplete.  Fortunately, estimates exist of the
contributions made by these terms to the nucleon's electromagnetic properties
\cite{oettel1,leinwebercohen,AWT1,AWT2}, and in the following they are used
to inform the model's application.

\section{Calculated Form Factors}
\label{sec:results}
The nucleon's electromagnetic current is
\begin{eqnarray}
\label{Jnucleon}
J_\mu(P^\prime,P) & = & ie\,\bar u(P^\prime)\, \Lambda_\mu(q,P) \,u(P)\,, \\
& = &  i e \,\bar u(P^\prime)\,\left( \gamma_\mu F_1(q^2) +
\frac{1}{2M}\, \sigma_{\mu\nu}\,q_\nu\,F_2(q^2)\right) u(P)\,,
\label{JnucleonB}
\end{eqnarray}
where $q= P^\prime - P$ and $\Lambda_\mu$ is the nucleon-photon vertex
described in the appendix.  In Eq.\ (\ref{JnucleonB}), $F_1$ and $F_2$ are,
respectively, the Dirac and Pauli electromagnetic form factors.  They are the
primary calculated quantities, from which one obtains the nucleon's electric
and magnetic form factors
\begin{equation}
\label{GEpeq}
G_E(q^2)  =  F_1(q^2) - \frac{q^2}{4 M^2} F_2(q^2)\,,\; 
G_M(q^2)  =  F_1(q^2) + F_2(q^2)\,.
\end{equation}

\begin{table}[t]
\beforetab
\begin{tabular}{c c c c c c c c}
\firsthline
\multicolumn{4}{c}{Parameters} &
\multicolumn{4}{c}{Calculated Static Properties} \\  \preline
$\mbox{\sc r}$ & $m_{0^+}$ (GeV) & $\omega_{0^+} $ & $\omega_{q\{qq\}}$ &
$(r_p^2)^2$ (fm$^2$) & 
$(r_n^2)^2$          & 
$\mu_p$ ($\mu_N$)    & 
$\mu_n$    \\\midhline
$0.25$ & $0.75$ & $1.50$ & $0.33$ & 
$(0.65)^2$ & $-(0.38)^2$ & $2.58$   & $-1.39$  \\ \preline 
$0.50$  & $0.77$ & $1.42$ & $0.29$ & 
$(0.61)^2$ & $-(0.37)^2$ & $2.52$   & $-1.37$  \\ \midhline
Obs.\       & & & & 
$(0.87)^2$ &$-(0.34)^2$ & $2.79$ & $-1.91$ \\
\lasthline
\end{tabular}
\aftertab \captionaftertab[]{\label{tablea} Fitted model parameters and
calculated nucleon static properties.  Experimental values are provided for
comparison in the last row.  NB.\ The values indicate:
$\lambda_{0^+}=1/m_{0^+} \sim \frac{1}{3}\,$fm, $l_{0^+}=1/\omega_{0^+}\sim
\frac{1}{7}\,$fm, $l_{q\{qq\}} = 1/\omega_{q\{qq\}} \sim \frac{2}{3}\,$fm.}
\end{table}

To proceed with the illustration we select two values of {\sc r}, namely,
$\mbox{\sc r} = 0.25$, $0.50$.  This choice is motivated by Faddeev equation
studies, in which the smaller value is obtained by calculations that retain
scalar and axial-vector diquark correlations but neglect the pion cloud,
while the latter is obtained if the estimated effect of that cloud is
incorporated \protect\cite{NpiN}.  Then, in each case, we fix the remaining
three parameters by requiring a least-squares fit to
\begin{equation}
\label{GEdipole}
G_{\!E}^p(q^2) = 1/(1+ q^2/m_{d}^2)^2\,, 
\end{equation}
with $m_{d}= 1.1\,$GeV.  This value of the dipole mass corresponds to a
proton charge radius $r_p=0.62\,$fm; i.e., $\sim 30\,$\% smaller than the
experimental value, and therefore leaves room deliberately for additional
contributions to the nucleon's electromagnetic structure from axial-vector
diquark correlations and a meson ``cloud.''  While the leading nonanalytic
contribution to $r_p$ can alone repair the discrepancy
\protect\cite{leinwebercohen,AWT2}, that does not alter the quiddity of our
scheme because the scale of the effect is clear and redistributing strength
between complementary contributions can be achieved merely by fine-tuning the
model's parameters.

\begin{figure}[t]
\vspace*{2.0ex}
\centerline{\includegraphics[height=0.75\textwidth]{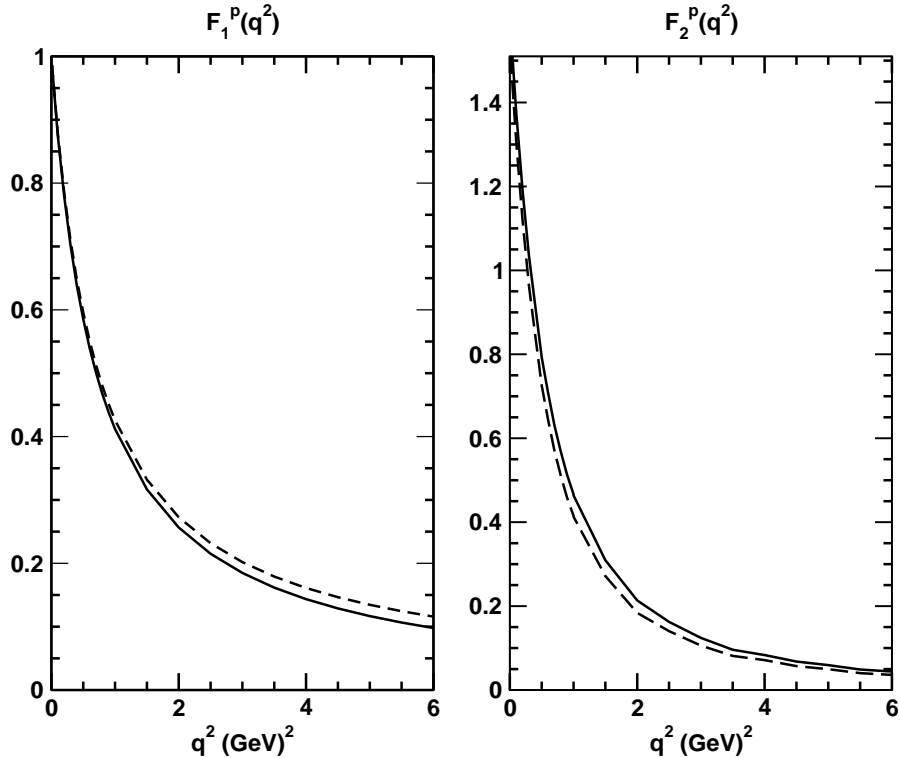}}
\caption{\label{calcF1p} Left panel -- Calculated proton Dirac form factor;
Right panel -- Calculated proton Pauli form factor.  Solid line, $\mbox{\sc
r}=0.25$; dashed, $\mbox{\sc r}=0.50$.  The associated model parameters are
listed in Table \protect\ref{tablea}.}
\end{figure}

In Table \ref{tablea} we list the parameter values produced by this fitting
procedure along with the calculated values of proton and neutron static
properties.  The proton's charge radius is precisely that for which we aimed.
However, the remaining values point to the deficiencies of a model that
retains only a scalar diquark correlation.
They confirm that in this case one is unable to obtain a quantitatively
accurate picture of the nucleon.
This is reassuring because, in contributing to nucleon observables,
axial-vector diquark correlations primarily interfere constructively with the
pion cloud; e.g., they both provide additional binding and hence act to
reduce the nucleon's mass \cite{NpiN}, and they both act to increase
$|\mu_{p,n}|$, $|r_{p,n}^2|$ \cite{oettel1,leinwebercohen,AWT1,AWT2}.  Hence
a model that ignores these contributions but succeeds in fitting experimental
data is likely to possess spurious degrees of freedom.\footnote{The
implementation of current conservation in the one-body current described in
the appendix is too simple to allow a fair description of $G_E^n$.  In this
case it misses important cancellations and hence we report anomalously large
values for $|r_{n}^2|$.  A realistic description requires a more complex
current which includes fully-fledged seagull terms \protect\cite{oettel1}.
The simple current is adequate for the remaining form factors because such
destructive interference is either absent or markedly less important
\protect\cite{bloch1}.}  The improvements necessary to make the model more
realistic are therefore plain.  In their absence it is nevertheless possible
to illustrate important points.

A first observation relates to what may be called the nucleon's ``quark
core.''  In an internally consistent model it is always possible to identify
the relative strength of various contributions to a physical observable.  For
example, Ref.\ \cite{NpiN} employs a rainbow-dressed quark and ladder-bound
meson basis, within which the quark core contributes approximately 85\% of
the nucleon's mass.  In the present case an estimate of the core's spatial
extent is afforded by $l_{q\{qq\}} = 1/\omega_{q\{qq\}} \approx
\frac{2}{3}\,$fm, which is commensurate with our calculated core contribution
to the proton's charge radius, Table \ref{tablea}.  It will also be evident
that this scale is consistent with estimates of the magnitude of meson-loop
contributions to the proton's charge radius determined from lattice-QCD
simulations \cite{leinwebercohen,AWT2}.

\begin{figure}[t]
\vspace*{7.5ex}

\centerline{\includegraphics[height=0.50\textwidth]{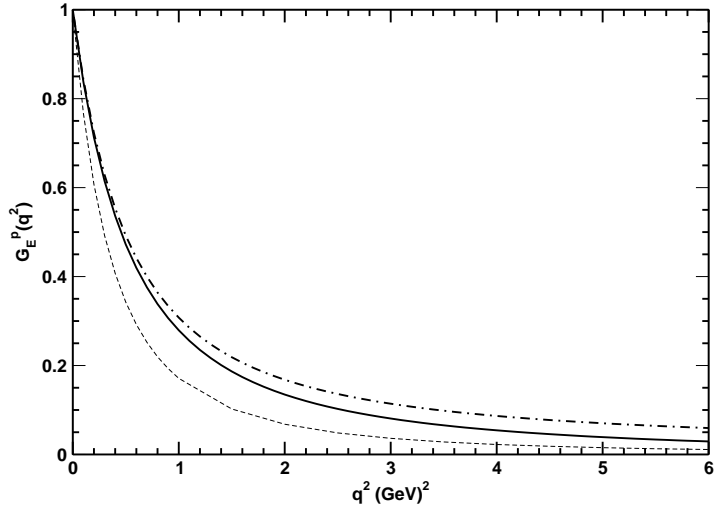}}
\caption{\label{calcGEp} Calculated proton electric form factor: solid line,
$\mbox{\sc r}=0.25$; dot-dashed, $\mbox{\sc r}=0.50$ -- the associated model
parameters are listed in Table \protect\ref{tablea}.  For comparison, the
lighter dashed line is $G_E^p(q^2)=1/(1+q^2/m_{\rm emp-p}^2)^2$, $m_{\rm
emp-p} = 0.84\,$GeV; viz., a dipole fit to available proton data.}
\end{figure}

\begin{figure}[t]
\vspace*{7.5ex}

\centerline{\includegraphics[height=0.50\textwidth]{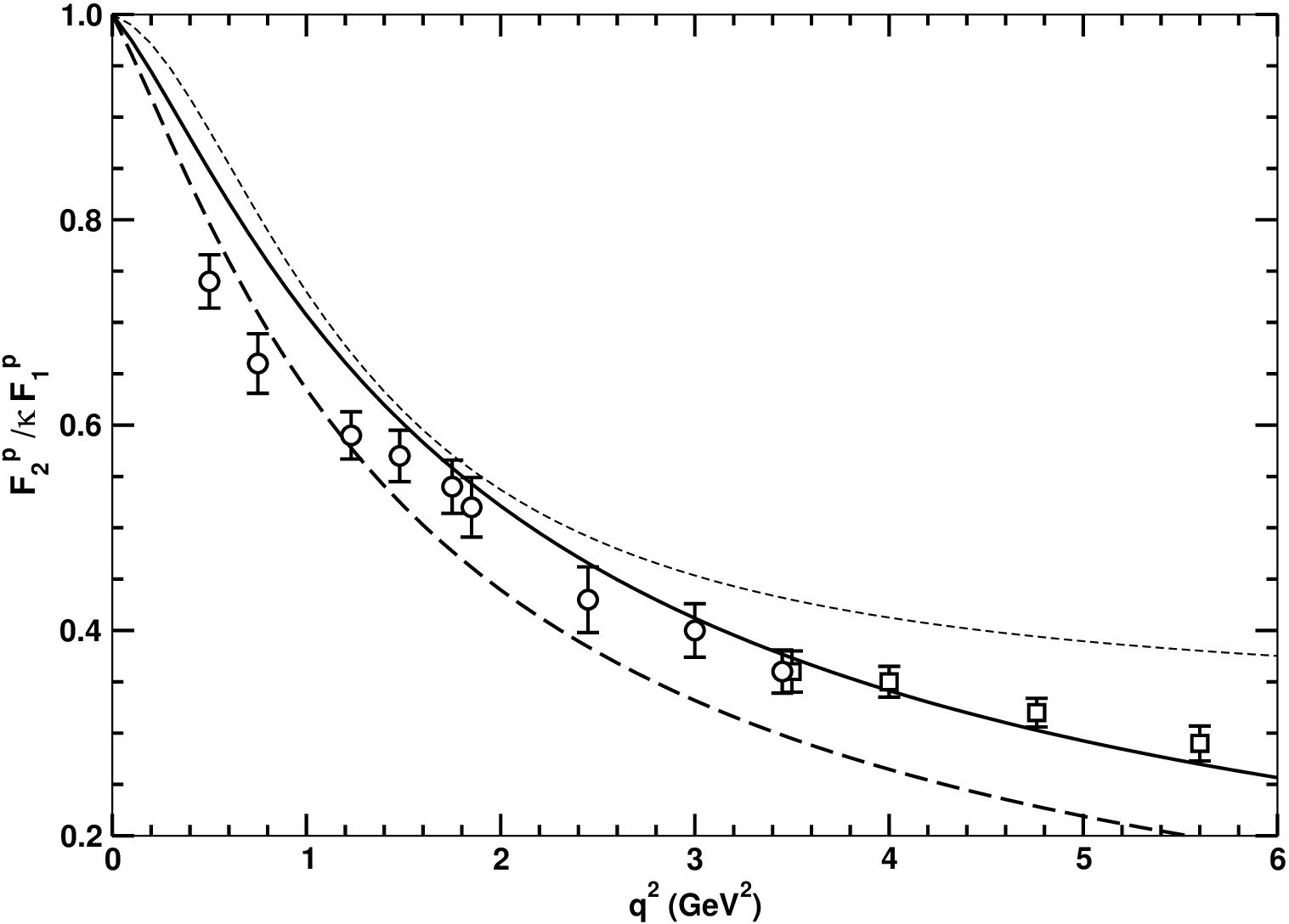}}
\caption{\label{calcF2F1} Calculated ratio $F_2^p/(\kappa F_1^p)$, $\kappa=
\mu_p-1$: solid line, $\mbox{\sc r}=0.25$; dashed, $\mbox{\sc r}=0.50$ -- the
associated model parameters are listed in Table \protect\ref{tablea}.  The
short-dashed line was obtained with the model parameters in Ref.\
\protect\cite{bloch1}: $\mbox{\sc r}=0.0$, and (in GeV) $m_{0^+}= 0.63$,
$\omega_{0^+} = 1.0$, $\omega_{q\{qq\}}= 0.2$.  Data: boxes,
Ref.\protect\cite{jones}; circles, Ref.\protect\cite{gayou}.}
\end{figure}

\begin{figure}[t]
\vspace*{7.5ex}

\centerline{\includegraphics[height=0.50\textwidth]{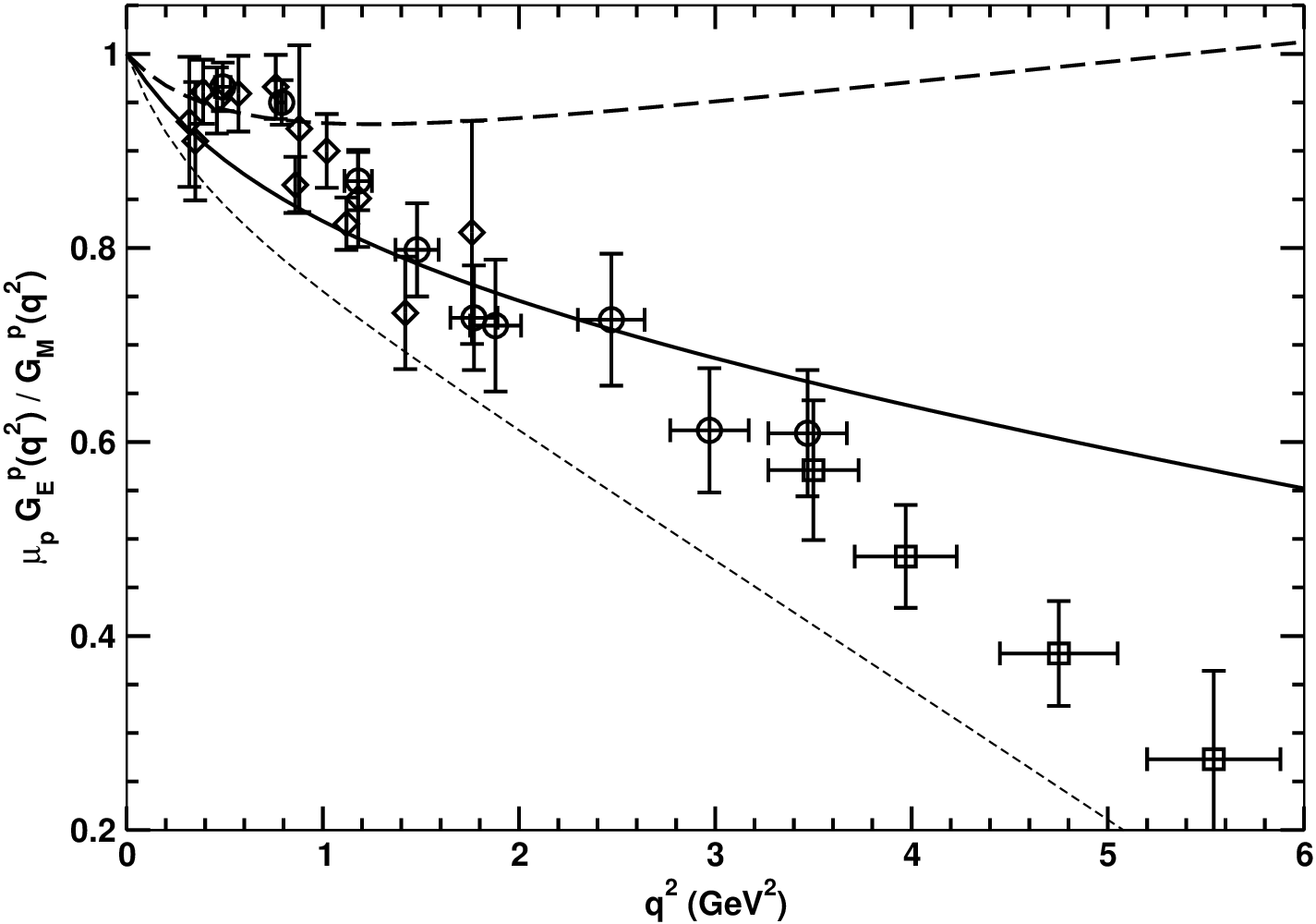}}
\caption{\label{calcpGEGM} Calculated ratio $\mu_p G_E^p(q^2)/G_M^p(q^2)$:
solid line, $\mbox{\sc r}=0.25$; dashed, $\mbox{\sc r}=0.50$ -- the
associated model parameters are listed in Table \protect\ref{tablea}. The
lighter short-dashed line was obtained with the model parameters in Ref.\
\protect\cite{bloch1}: $\mbox{\sc r}=0.0$, and (in GeV) $m_{0^+}= 0.63$,
$\omega_{0^+} = 1.0$, $\omega_{q\{qq\}}= 0.2$.  Data: boxes,
Ref.\protect\cite{jones}; diamonds, Ref.\protect\cite{roygayou}; circles,
Ref.\protect\cite{gayou}.}
\end{figure}

We depict the proton's Dirac form factor in the left panel of Fig.\
\ref{calcF1p}, wherein it is clear that the shift in parameter values has
little observable impact.  That is also true for the Pauli form factor, which
is plotted in the right panel.  These two functions together give the
proton's electric form factor, via Eq.\ (\ref{GEpeq}), and that is shown in
Fig.\ \ref{calcGEp}.

We plot the calculated ratio: $F_2/\kappa \, F_1$, $\kappa = \mu_p-1$, in
Fig.\ \ref{calcF2F1}, along with modern experimental data \cite{jones,gayou}.
In addition, we draw the calculated result for this ratio obtained using the
parameter values of the scalar diquark \textit{Ansatz} employed in Ref.\
\cite{bloch1}.  These values: $\mbox{\sc r}=0.0$, and (in GeV) $m_{0^+}=
0.63$, $\omega_{0^+} = 1.4$, $\omega_{q\{qq\}}= 0.2$, were fixed via a
least-squares fit to Eq.\ (\ref{GEdipole}) but with $m_{d}=0.84\,$GeV, which
gives $r_p=0.87\,$fm.  The figure illustrates that the ratio decreases
smoothly with increasing $\mbox{\sc r}$, bracketing the data, and thereby
suggests that even the rudimentary \textit{Ansatz} is capable of accurately
describing the data.  Indeed, the $\mbox{\sc r}=0.25$ parameter set might
even be considered a good representation.  However, this is not the
conclusion we draw.  Rather, the result demonstrates that the available data
on this ratio is sensitive to model-dependent details.

The ratio $\mu_p G_E^p/G_M^p$ is depicted in Fig.\ \ref{calcpGEGM} and
therein the last statement is amplified.  The proton's electric form factor
is a difference, Eq.\ (\ref{GEpeq}), and that accentuates its sensitivity.
On the domain for which data is available, this ratio is particularly
responsive to model details, a result also conspicuous in Ref.\
\cite{oettel1}.  It is plain that three parameter sets, which are reasonable
and differ modestly when compared through the ratio $F_2/\kappa \, F_1$,
appear vastly different in the comparison presented in Fig. \ref{calcpGEGM}.
Moreover, because of continuity, it is clear that one could tune the model
parameters to fit the data on this ratio.  However, what might be considered
success in that endeavour could easily be achieved through results for
$G_E^p$ and $G_M^p$ individually which both disagree with the data.

\section{Summary and Discussion}
\label{sec:epilogue}
We calculated the proton's elastic electromagnetic form factors using a
rudimentary \textit{Ansatz} for the nucleon's Poincar\'e covariant Faddeev
amplitude that represents the proton as a composite of a confined quark and
confined nonpointlike scalar diquark.  All such models give a Faddeev wave
function that corresponds to a nucleon spinor with a sizeable lower component
in the rest frame.

This study indicates that on the domain of $q^2$ accessible in modern
precision experiments these form factors are a sensitive probe of
nonperturbative dynamics.  The calculated pointwise forms express a
dependence on the length-scales that characterise nonperturbative phenomena,
such as bound state extent, dressing of quark and gluon propagators, and
meson cloud effects.  This is precisely analogous to the current status of
the pion's electromagnetic form factor, for which the behaviour predicted in
a straightforward application of perturbative QCD is not unambiguously
evident until $q^2 \gtrsim 15\,$GeV$^2$ \cite{mrpion}.
  
The ratio $\mu_p G_E^p/G_M^p$ is particularly sensitive to infrared dynamics
because, as $q^2$ increases, $G_E^p(q^2)$ is a difference of small
quantities.  However, this ratio should not be considered in isolation
because it is possible to reproduce the experimental data using a model that
simultaneously provides a poor description of the individual form factors.
That is also true of the ratio $F_2^p/\kappa F_1^p$ but this combination is
less responsive to model particulars because the Dirac and Pauli form factors
are positive and fall uniformly to zero, with a momentum dependence at
asymptotically large momenta given by perturbative QCD
\cite{brodskylepage1,brodskylepage2}.  In the absence of a veracious
theoretical understanding of the nucleon, we view the latter ratio as a more
sensible constraint on contemporary studies.

It is apparent that much can be learnt about long-range dynamics in QCD from
existing and forthcoming accurate data on nucleon form factors.  In
constraining systematic QCD-based calculations, one can hope, for example, to
see an evolution from the domain on which meson cloud effects are important
to that whereupon observables are dominated by the confined quark core.  This
could be elucidated by improving the present study; viz., basing it on a
solution of the Poincar\'e covariant Faddeev equation with axial-vector
diquark correlations, instead of using an \textit{Ansatz} for the Faddeev
amplitude, and explicitly including meson cloud contributions.  The form
factors would then be tied directly to assumptions about the nature of
quark-gluon dynamics in the nucleon. 

\bigskip

\begin{acknowledge}
We are grateful to M.\,B.\ Hecht for helpful correspondence and for providing
us with his computer code; and to F.\ Coester and T.-S.\,H.\ Lee for
constructive discussions.  This work was supported by: the Department of
Energy, Nuclear Physics Division, under contract no.\ \mbox{W-31-109-ENG-38};
the Austrian Research Foundation FWF under Erwin-Schr\"odinger-Stipendium
no.\ J2233-N08; and benefited from the resources of the National Energy
Research Scientific Computing Center.
\end{acknowledge}

\bigskip

\appendix
\section{Nucleon-Photon Vertex}
\begin{figure}[t]
\centerline{\includegraphics[height=0.18\textwidth]{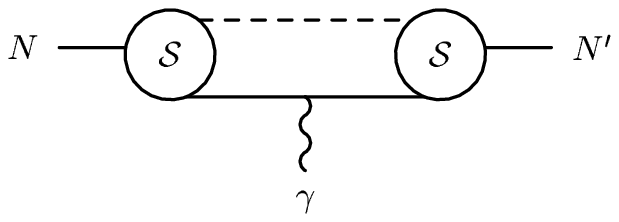}%
\includegraphics[height=0.18\textwidth]{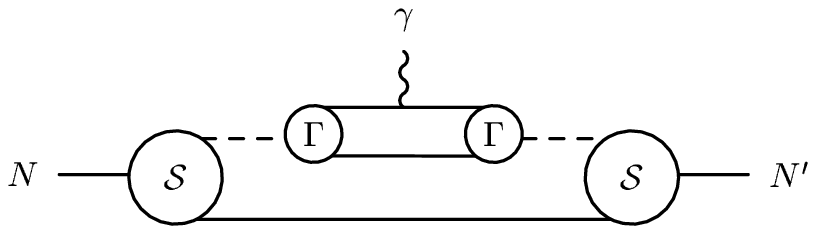}}

\centerline{\includegraphics[height=0.18\textwidth]{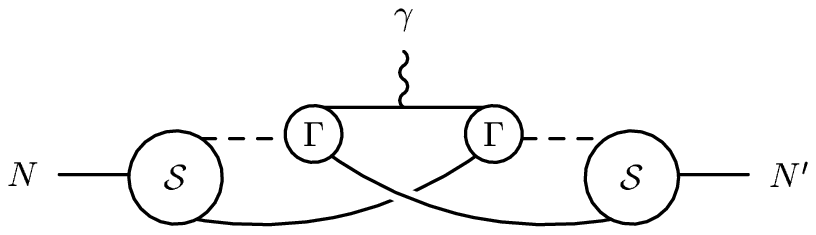}}
\vspace*{2ex}

\centerline{\includegraphics[height=0.17\textwidth]{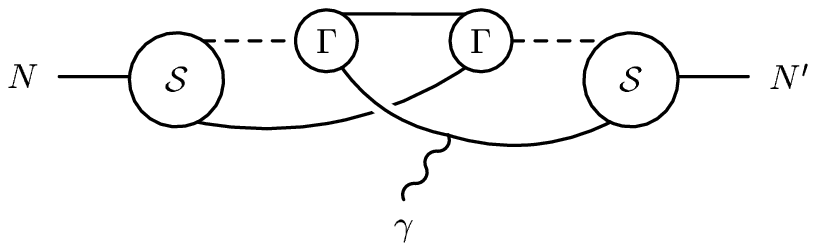}%
\includegraphics[height=0.17\textwidth]{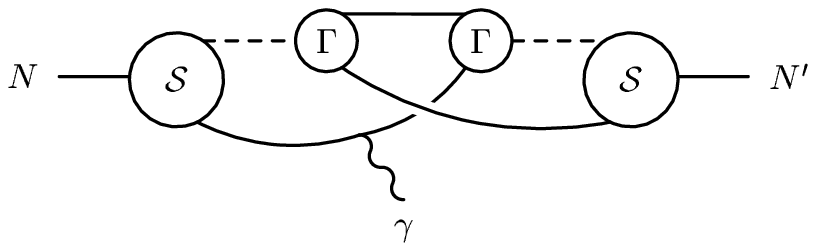}}
\caption{One-body current obtained from the product \textit{Ansatz} of Eq.\
(\protect\ref{calS}).  Solid line, dressed-quark propagator $S(p)$; dashed
line, diquark propagator $\Delta^{0^+}(K)$.  The amputated nucleon vertex is
$\Psi_3^{0^+}\!\!$ in every case.
\label{onebody}}
\end{figure}
We use an \textit{Ansatz} for $\Psi_3$ in the nucleon's Faddeev amplitude,
Eq.\ (\ref{PsiT}), from which a properly antisymmetrised one-body vertex can
be constructed via the method outlined in Ref.\ \cite{revbasti}:
\begin{eqnarray}
\label{nucvtx}
\Lambda_\mu(q,P) &=&
\Lambda^1_\mu(q,P)
+ 2 \sum_{i=2}^5\,\Lambda^i_\mu(q,P)\,,
\end{eqnarray}
wherein:
\begin{equation}
\Lambda^1_\mu(q,P)  =  3 \int\!\!\sfrac{d^4 \ell}{(2\pi)^4}\,
{\cal S}(\ell- \sfrac{2}{3}q;P)
\,\Delta^{0^+}(K)\,{\cal S}(\ell;P)\,\Lambda_\mu^S(p_3+q,p_3) \,, 
\label{L1}
\end{equation}
with $K=\ell + \sfrac{2}{3} P$, $p_3= \sfrac{1}{3} P - \ell$,
$\Lambda_\mu^S(\ell_1,\ell_2) =
S(\ell_1)\,\Gamma_\mu^Q(\ell_1,\ell_2)\,S(\ell_2)$; and
\begin{equation}
\Lambda^2_\mu(q,P)  =  6\int\!\!\sfrac{d^4 k}{(2\pi)^4}\sfrac{d^4
\ell}{(2\pi)^4}\, \Omega(p_1+q,p_2,p_3)\,\Omega(p_1,p_2,p_3) \, {\rm
tr}_{DF}\left[\Lambda_\mu^S(p_1+q,p_1) S(p_2)\right]\, S(p_3)\,,
\end{equation}
where $p_1= \sfrac{1}{2} K + k$, $p_2= \sfrac{1}{2} K - k$, $6 =
\varepsilon_{c_1 c_2 c_3} \varepsilon_{c_1 c_2 c_3}$ is the colour
contraction, and
\begin{equation}
\Omega(p_1,p_2,p_3) = \Delta^{0^+}(p_{\{12\}}) \,
\Gamma_{0^+}(\sfrac{1}{2}p_{[12]};p_{\{12\}})\, {\cal
S}(\sfrac{1}{3}[p_{\{12\}}-2p_3];P) \,.
\end{equation}
$\Lambda_\mu^2$ describes the photon probing the structure of the scalar
diquark correlation, and contributes equally to both the proton and neutron.
The remaining terms are
\begin{eqnarray}
\nonumber\lefteqn{\Lambda^3_\mu(q,P) = 6\int\!\!\sfrac{d^4
k}{(2\pi)^4}\sfrac{d^4 \ell}{(2\pi)^4}\, \Omega(p_1,p_2,p_3)\, }\\
&& \times\, \Omega(p_1+q,p_3,p_2)\,S(p_2) \, (i\tau_2)^{\rm t}
\Lambda^S_\mu(p_1,p_1+q)\,(i \tau_2)\,S(p_3)\,,
\label{Lambda3}
\end{eqnarray}
where ``t'' denotes matrix transpose, and 
\begin{eqnarray}
\nonumber\Lambda^4_\mu(q,P) & = & 6\int\sfrac{d^4 k}{(2\pi)^4} \sfrac{d^4
\ell}{(2\pi)^4}\,\Omega(p_1,p_3,p_2+q)\, \Omega(p_1,p_2,p_3)
\,\Lambda^S_\mu(p_2+q,p_2) \,S(p_1) \, S(p_3)\,,\\
&&\label{L4} \\
\nonumber \Lambda^5_\mu(q,P) & = & 6\int\sfrac{d^4 k}{(2\pi)^4}\sfrac{d^4
\ell}{(2\pi)^4}\, \Omega(p_1,p_3+q,p_2)\Omega(p_1,p_2,p_3)\,
S(p_2)\,S(p_1)\,\Lambda_\mu^S(p_3+q,p_3) \,.\\
\label{L5}
\end{eqnarray}
We illustrate these five terms in Fig.\ \ref{onebody}.  They are in
one-to-one correspondence with those considered in Ref.\
\cite{OettelPichowsky}, with the bottom two diagrams, representing
$\Lambda_\mu^{4,5}$, being progenitors of the ``seagull'' terms exploited
therein to ensure current conservation.  

Our results are obtained by evaluating these integrals using Monte-Carlo
methods and the input specified in Eqs.\ (\ref{ssm}), (\ref{svm}),
(\ref{Delta0p}), (\ref{Gamma0p}), (\ref{SAnsatz}) and (\ref{bcvtx}).


\begin{thebibliography}{99}

\bibitem{latticepion1} G.\ Martinelli and C.\,T.\ Sachrajda,
Nucl.\ Phys.\ B {\bf 306}, 865 (1988).

\bibitem{draper} T.\ Draper, R.\,M.\ Woloshyn, W.\ Wilcox and K.\,F.\ Liu,
Nucl.\ Phys.\ B {\bf 318}, 319 (1989).

\bibitem{laermannpion} J.\ van der Heide, M.\ Lutterot, J.\,H.\ Koch and E.\
Laermann, ``The pion form factor in improved lattice QCD,''
hep-lat/0303006.

\bibitem{revbasti} C.\,D.\ Roberts and S.\,M.\ Schmidt, Prog.\ Part.\ Nucl.\
Phys.\ \textbf{45}, S1 (2000).

\bibitem{revreinhard} R.\ Alkofer and L.\,v.\ Smekal, 
Phys.\ Rept.\ {\bf 353}, 281 (2001). 

\bibitem{revpieter} P.\ Maris and C.\,D.\ Roberts, ``Dyson-Schwinger
equations: A tool for hadron physics,'' nucl-th/0301049, to appear in
Int. J. Mod. Phys. E.

\bibitem{pieterpion} P.\ Maris and P.\,C.\ Tandy, 
Phys.\ Rev.\ {\bf C 61}, 045202 (2000). 
 
\bibitem{latticeform1} G.\ Martinelli and C.\,T.\ Sachrajda,
Nucl.\ Phys.\ B {\bf 316}, 355 (1989).

\bibitem{latticeform} M.\ G\"ockeler, \textit{et al}., 
``Nucleon electromagnetic form factors on the lattice and in chiral effective
field theory,'' hep-lat/0303019.

\bibitem{bloch1} J.\,C.\,R.\ Bloch, C.\,D.\ Roberts, S.\,M.\ Schmidt, A.\
Bender and M.\,R.\ Frank,
Phys.\ Rev.\ C {\bf 60}, 062201 (1999).

\bibitem{bloch2} J.\,C.\,R.\ Bloch, C.\,D.\ Roberts and S.\,M.\ Schmidt,
Phys.\ Rev.\ C {\bf 61}, 065207 (2000).

\bibitem{oettel1} M.\ Oettel, R.\ Alkofer and L.\,v.\ Smekal,
Eur.\ Phys.\ J.\ A {\bf 8}, 553 (2000).

\bibitem{coester} P.\,L.\ Chung and F.\ Coester,
Phys.\ Rev.\ D {\bf 44}, 229 (1991).

\bibitem{revcoester} F.\ Coester,
Prog.\ Part.\ Nucl.\ Phys.\  {\bf 29}, 1 (1992).

\bibitem{salme} E.\ Pace, G.\ Salme and A.\ Molochkov,
Nucl.\ Phys.\ A {\bf 699}, 156 (2002).

\bibitem{plessas} S.\ Boffi, \textit{et al}., 
Eur.\ Phys.\ J.\ A {\bf 14}, 17 (2002).

\bibitem{millerfrank} G.\,A.\ Miller and M.\,R.\ Frank,
Phys.\ Rev.\ C {\bf 65}, 065205 (2002).

\bibitem{justindq} R.\,T.\ Cahill, C.\,D.\ Roberts and J.\ Praschifka,
Phys.\ Rev.\ D {\bf 36}, 2804 (1987).

\bibitem{pieterdq} P.\ Maris,
Few Body Syst.\ {\bf 32}, 41 (2002).

\bibitem{latticedq} M.\ Hess, F.\ Karsch, E.\ Laermann and I.\ Wetzorke,
Phys.\ Rev.\ D {\bf 58}, 111502 (1998).

\bibitem{truncscheme} A.\ Bender, C.\,D.\ Roberts and L.\,v.\ Smekal,
Phys.\ Lett.\ B {\bf 380}, 7 (1996).

\bibitem{reinharddq} G.\ Hellstern, R.\ Alkofer and H.\ Reinhardt,
Nucl.\ Phys.\ A {\bf 625}, 697 (1997).

\bibitem{detmold} A.\ Bender, W.\ Detmold, C.\,D.\ Roberts and A.\, W.\
Thomas,
Phys.\ Rev.\ C {\bf 65}, 065203 (2002).

\bibitem{regfe} R.\,T.\ Cahill, C.\,D.\ Roberts and J.\ Praschifka, Austral.\
J.\ Phys.\ {\bf 42}, 129 (1989). 

\bibitem{NpiN} M.\,B.\ Hecht, M.\ Oettel, C.\,D.\ Roberts, S.\,M.\ Schmidt,
P.\,C.\ Tandy and A.\,W.\ Thomas,
Phys.\ Rev.\ C {\bf 65}, 055204 (2002).

\bibitem{oettel2} M.\ Oettel, G.\ Hellstern, R.\ Alkofer and H.\ Reinhardt, 
Phys.\ Rev.\ C {\bf 58}, 2459 (1998).

\bibitem{brash} E.\,J.\ Brash, A.\ Kozlov, S.\ Li and G.\,M.\ Huber,
Phys.\ Rev.\ C {\bf 65}, 051001 (2002).

\bibitem{arrington} J.\ Arrington, ``How well do we know the electromagnetic
form factors of the proton?,'' nucl-ex/0305009.

\bibitem{jones} M.\,K.\ Jones {\it et al.}  [JLab Hall A Collaboration],
Phys.\ Rev.\ Lett.\ {\bf 84}, 1398 (2000).

\bibitem{roygayou} O.\ Gayou, {\it et al.},
Phys.\ Rev.\ C {\bf 64}, 038202 (2001).

\bibitem{gayou} O.\ Gayou, {\it et al.}  [JLab Hall A Collaboration],
Phys.\ Rev.\ Lett.\ {\bf 88}, 092301 (2002). 

\bibitem{ralston} T.\ Gousset, B.\ Pire and J.\,P.\ Ralston,
Phys.\ Rev.\ D {\bf 53}, 1202 (1996).

\bibitem{mrt98} P.\ Maris, C.\,D.\ Roberts and P.\,C.\ Tandy,
Phys.\ Lett.\ B {\bf 420}, 267 (1998).

\bibitem{jackson} G.\,R.\ Farrar and D.\,R.\ Jackson,
Phys.\ Rev.\ Lett.\  {\bf 43}, 246 (1979).

\bibitem{mrpion} P.\ Maris and C.\,D.\ Roberts,
Phys.\ Rev.\ C {\bf 58}, 3659 (1998).

\bibitem{bowman} P.\,O.\ Bowman, U.\,M.\ Heller and A.\,G.\ Williams,
Phys.\ Rev.\ D {\bf 66}, 014505 (2002).

\bibitem{mandar} M.\,S.\ Bhagwat, M.\,A.\ Pichowsky, C.\,D.\ Roberts and
P.\,C.\ Tandy, ``Analysis of a quenched lattice-QCD dressed-quark
propagator,'' nucl-th/0304003, to appear in Phys.\ Rev.\ C.

\bibitem{mr97} P.\ Maris and C.\,D.\ Roberts, Phys.\ Rev.\ C {\bf 56}, 3369
(1997).

\bibitem{mark} C.\,J.\ Burden, C.\,D.\ Roberts and M.\,J.\ Thomson, Phys.\
Lett.\ {\bf B 371}, 163 (1996).

\bibitem{simon} S.\ Capstick and W.\ Roberts, Prog.\ Part.\ Nucl.\ Phys.\
{\bf 45}, S241 (2000).

\bibitem{nedm} M.\,B.\ Hecht, C.\,D.\ Roberts and S.\,M.\ Schmidt,
Phys.\ Rev.\ C {\bf 64}, 025204 (2001).
 
\bibitem{anomalyBando} M.\ Bando, M.\ Harada and T.\ Kugo, Prog.\ Theor.\
Phys.\ {\bf 91}, 927 (1994).

\bibitem{anomalyRoberts1} C.\,D.\ Roberts,
Nucl.\ Phys.\ {\bf A 605}, 475 (1996).

\bibitem{anomalyAlkofer} R.\ Alkofer and C.\,D.\ Roberts, Phys.\ Lett.\ {\bf
B 369}, 101 (1996).

\bibitem{anomalyKlabucar} D.\ Kekez and D.\ Kla\-bu\-\v{c}ar, Phys.\ Lett.\
{\bf B 457}, 359 (1999).

\bibitem{anomalyRoberts2} C.\,D.\ Roberts,
Fizika {\bf B 8}, 285 (1999).

\bibitem{anomalyTandy} P.\,C.\ Tandy, Fizika {\bf B 8}, 295 (1999).

\bibitem{ayse1} A.\ Bashir, A.\ K\i z\i lers\"u and M.\,R.\ Pennington,
Phys.\ Rev.\ D {\bf 57}, 1242 (1998).

\bibitem{ayse2} A.\ Bashir, A.\ K\i z\i lers\"u and M.\,R.\ Pennington,
``Analytic form of the one-loop vertex and of the two-loop fermion propagator
in 3-dimensional massless QED,'' hep-ph/9907418.

\bibitem{Volmer:2000ek} J.\ Volmer, {\it et al.}  [JLab $F_\pi$
Collaboration],
Phys.\ Rev.\ Lett.\ {\bf 86}, 1713 (2001). \label{RVolmer:2000ek}
 
\bibitem{bc80} J.\,S.\ Ball and T.-W.\ Chiu, Phys.\ Rev.\ D {\bf 22}, 2542
(1980).

\bibitem{pichowsky} M.\,A.\ Pichowsky, S.\ Walawalkar and S.\ Capstick, Phys.\
Rev.\ D {\bf 60}, 054030 (1999).

\bibitem{piloop} R.\ Alkofer, A.\ Bender and C.\,D.\ Roberts,
Int.\ J.\ Mod.\ Phys.\ {\bf A 10}, 3319 (1995).

\bibitem{salme2} F.\ Cardarelli, E.\ Pace, G.\ Salme and S.\ Simula, 
Phys.\ Lett.\ B {\bf 357}, 267 (1995).

\bibitem{leinwebercohen} D.\,B.\ Leinweber and T.\,D.\ Cohen,
Phys.\ Rev.\ D {\bf 47}, 2147 (1993).

\bibitem{AWT1} E.\,J.\ Hackett-Jones, D.\,B.\ Leinweber and A.\,W.\ Tho\-mas,  
Phys.\ Lett.\ {\bf B 489}, 143 (2000).

\bibitem{AWT2} E.\,J.\ Hackett-Jones, D.\,B.\ Leinweber and A.\,W.\ Tho\-mas, 
Phys.\ Lett.\ {\bf B 494}, 89 (2000).

\bibitem{OettelPichowsky} M.\ Oettel, M.\,A.\ Pichowsky and L.\,v.\ Smekal,
Eur.\ Phys.\ J.\ A {\bf 8}, 251 (2000).

\bibitem{brodskylepage1} G.\,P.\ Lepage and S.\,J.\ Brodsky,
Phys.\ Rev.\ Lett.\ {\bf 43}, 545 (1979) [Erratum-ibid.\ {\bf 43}, 1625
(1979)].

\bibitem{brodskylepage2} G.\,P.\ Lepage and S.\,J.\ Brodsky,
Phys.\ Rev.\ D {\bf 22}, 2157 (1980).

\end{thebibliography}
\end{document}